\documentclass[a4paper,11pt]{article}
\usepackage{jheppub}
\usepackage{eurosym}
\usepackage{amssymb}
\usepackage{amsfonts,cite}
\usepackage{amsmath}
\usepackage{graphicx}
\usepackage{multirow}
\usepackage{xcolor}
\usepackage{epsfig}
\usepackage{fancyhdr}

\newbox\mybox

\newcommand\fverb{\setbox\mybox=\hbox\bgroup\verb}
\newcommand\fverbdo{\egroup\medskip\noindent\fbox{\unhbox\mybox}\ }
\newcommand\fverbit{\egroup\item[\fbox{\unhbox\mybox}]}

\abstract{
		We develop a Bohmian analysis of a two-dimensional ghost Hamiltonian and its mapping to the degenerate Pais-Uhlenbeck model. Using Gaussian wavepackets, we derive the corresponding guidance equations, the centre and width evolution, and the quantum potential. We use these quantities to characterise bounded, quasi-semiclassical, spiral, and runaway regimes. The Bohmian trajectories provide a direct dynamical diagnostic of coherence, packet deformation, and quantum-classical separation. We then compare a bi-Hamiltonian pair consisting of the ghost Hamiltonian and a classically equivalent alternative formulation. While the two descriptions produce identical classical trajectories, they lead to different Bohmian trajectories and different quantum potentials evaluated along those trajectories. This demonstrates that classical equivalence need not extend to Bohmian quantum dynamics and identifies a concrete quantum ambiguity in the degenerate higher-derivative system.
}

\author[a]{Sanjib Dey and }
\author[b]{Andreas Fring}

\title{Quantum-classical diagnostics and Bohmian inequivalence for higher time-derivative Hamiltonians}

\affiliation[a]{Department of Physics, BITS-Pilani, KK Birla Goa Campus, Zuarinagar, Goa 403726, India}
\affiliation[b]{Department of Mathematics, City St George's, University of London,  Northampton Square, \\ London EC1V 0HB, UK}
	
	\emailAdd{sanjibd@goa.bits-pilani.ac.in}
	\emailAdd{a.fring@city.ac.uk}

\begin{document}
	\maketitle

\pagestyle{fancy}
\fancyhead{} 
\fancyhead[LE,RO]{\small\itshape  Quantum-classical diagnostics and Bohmian inequivalence for HTD Hamiltonians} 

\renewcommand{\headrulewidth}{0.4pt}
	
\section{Introduction}	

Higher time-derivative theories (HTDTs) arise naturally in a wide range of physical contexts, including effective field theory  \cite{buchbinder2017effective}, modified gravity \cite{stelle77ren,starobinsky1980new,sotiriou2010f}, regularisation schemes \cite{stelle77ren,grav3}, and attempts at ultraviolet completion \cite{barnaby2008dynamics,tomboulis2015renormalization,modesto2016superrenormalizable}. Their main structural difficulty is well known: in non-degenerate settings, higher derivatives generically lead to Ostrogradsky instabilities \cite{ostrogradsky1850memoire,ghostconst,motohashi1,motohashi4,Woodard1}, which manifest themselves classically through unbounded Hamiltonians and quantum mechanically through ghost sectors, unbounded spectra, or non-normalisable states \cite{fring2025quant,FFT}. Determining whether such systems admit a meaningful quantum interpretation, and in what sense different Hamiltonian formulations should be regarded as equivalent, remains a central open question.

A particularly useful laboratory for these issues is provided by the Pais-Uhlenbeck (PU) oscillator \cite{pais1950field} and its equivalent ghost Hamiltonian formulations \cite{fring2025quant,FFT}. In these systems one encounters, already at the level of exactly solvable models, the characteristic tension between bounded spectra and normalisability, as well as the possibility of different Hamiltonian descriptions generating the same classical equations of motion \cite{fring2025quant,FFT}. This makes them especially suitable for probing the distinction between classical equivalence and quantum equivalence. In the degenerate case, where the two PU frequencies coincide, the dynamics becomes particularly subtle: the classical motion develops Jordan-block features and runaway secular terms, while the quantum theory exhibits additional ambiguities and a nontrivial algebraic structure with hidden symmetries \cite{fring2026spec}.

In this work we analyse these systems from the viewpoint of Bohmian mechanics \cite{Bohm1952,Holland1993,durr2009bohmian,sanz2012trajectory,deyfringBohmpra}. Rather than focusing only on spectral properties, we study the actual trajectory flow generated by the wavefunction. This has two advantages. First, it provides a direct dynamical diagnostic of the extent to which a given wavepacket behaves semiclassically, remains coherent, or develops strong quantum distortions. Second, it allows one to compare classically equivalent Hamiltonian formulations at the level of their pilot-wave dynamics, thereby revealing quantum differences that are not visible in the classical equations alone.

We begin with a two-dimensional ghost Hamiltonian and construct Gaussian wavepackets whose centres follow the classical motion while their widths evolve according to a Riccati equation. The corresponding Bohmian trajectories are then compared with classical trajectories and with the motion of the Gaussian centre. This leads to a set of trajectory-based diagnostics, including the internal deviation from the centre, the quantum-classical separation, and the quantum potential evaluated along the Bohmian paths. These diagnostics allow us to distinguish several qualitatively different regimes, ranging from rigid transport and quasi-semiclassical motion to unstable spiral behaviour, critical runaway motion, and non-normalisable Gaussian sectors.

A second objective is to connect these diagnostics to HTDTs more directly. The ghost Hamiltonian considered here is related to a higher time-derivative model through the standard reduction to first-order form. This allows us not merely to reinterpret the ghost dynamics, but to formulate a corresponding Bohmian diagnostic framework for the associated higher time-derivative system.

The most striking effect appears in the bi-Hamiltonian setting. In addition to the ghost Hamiltonian \(H_g\), the same degenerate classical dynamics admits a second Hamiltonian description \(H_2\) with a different kinetic structure \cite{FFT,fring2026spec}. Although the two Hamiltonians generate identical classical trajectories, we find that their Bohmian trajectories and trajectory-evaluated quantum potentials differ. This provides a concrete example of a quantum ambiguity hidden behind classical equivalence: the classical vector field is the same, but the pilot-wave dynamics is not.

Our paper is organised as follows: In section 2 we formulate the Bohmian dynamics of the two-dimensional ghost Hamiltonian and develop its interpretation in the corresponding higher time-derivative setting obtained by reduction to first-order form.
 In section 3 we analyse the Gaussian wavepacket dynamics and use Bohmian diagnostics to classify several dynamical regimes. In section 4 we turn to the bi-Hamiltonian structure and show that classically equivalent Hamiltonians can lead to inequivalent Bohmian flows. Section 5 contains our conclusions.

\section{Bohmian dynamics for a 2D ghost Hamiltonian and HTDTs}

\subsection{From the Schrödinger equation to quantum potentials}

We consider a two-dimensional quantum system with configuration variables
\(
q=(x,y)
\)
and Hamiltonian operator of the form
\begin{equation}
	\hat H
	=
	\frac{1}{2}\,\hat p_i\,G^{ij}\,\hat p_j
	+
	V(x,y),
	\qquad
	i,j\in\{x,y\},
\end{equation}
where \(G^{ij}\) is a constant symmetric matrix, not necessarily positive definite.
In the coordinate representation \(\hat p_i=-i\hbar\partial_i\), the Schrödinger equation reads
\begin{equation}
	i\hbar\,\partial_t\psi(x,y,t)
	=
	\left[
	-\frac{\hbar^2}{2}\,
	\partial_i\!\left(G^{ij}\partial_j\right)
	+
	V(x,y)
	\right]\psi(x,y,t).
	\label{Schroedinger}
\end{equation}
We write the wavefunction in polar (WKB) form
\begin{equation}
	\psi(x,y,t)=R(x,y,t)\,e^{iS(x,y,t)/\hbar},
	\qquad
	R\ge0, \label{WKB}
\end{equation}
with real amplitude \(R\) and real phase \(S\). Substituting this Ansatz into equation \eqref{Schroedinger} and separating real and imaginary parts yields two equations.
The imaginary part gives the continuity equation
\begin{equation}
	\partial_t(R^2)
	+
	\partial_i\!\left(
	R^2\,G^{ij}\,\partial_j S
	\right)
	=0.
	\label{continuity}
\end{equation}
This equation is the continuity equation for the density $\rho$ with associated current $j$,
\begin{equation}
	\rho=|\psi|^2=R^2,
	\qquad \text{and} \qquad
	j^i=
	R^2\,G^{ij}\,\partial_j S.
	\label{current}
\end{equation}
In turn, the real part yields the quantum Hamilton-Jacobi equation
\begin{equation}
	\partial_t S
	+
	\frac{1}{2}\,
	\partial_i S\,G^{ij}\,\partial_j S
	+
	V
	+
	Q
	=0,
	\label{QHJ}
\end{equation}
where the quantum potential is identified as
\begin{equation}
	Q
	=
	-\frac{\hbar^2}{2}\,
	\frac{1}{R}\,
	\partial_i\!\left(G^{ij}\partial_j R\right).
	\label{Qpotential}
\end{equation}
Equations \eqref{continuity} and \eqref{QHJ} are completely equivalent to the Schrödinger equation \eqref{Schroedinger}. No further assumptions have been made so far.

\subsection{Bohmian guidance law from equivariance}

In Bohmian mechanics one postulates that the configuration variables
\(q(t)=(x(t),y(t))\) follow deterministic trajectories guided by the wavefunction  \cite{Bohm1952,Holland1993,durr2009bohmian,sanz2012trajectory}.
Denoting the velocity field by \(v(q,t)\), an ensemble of such trajectories with density \(\rho(q,t)\) satisfies the transport equation
\begin{equation}
	\partial_t\rho
	+
	\partial_i(\rho\,v^i)
	=0.
	\label{transport}
\end{equation}
Requiring equivariance, i.e.\ that \(\rho(q,t)=|\psi(q,t)|^2=R^2(q,t)\) at all times, the transport equation \eqref{transport} must coincide with the continuity equation \eqref{continuity}.
Comparing the two equations, we identify
\begin{equation}
	\rho\,v^i
	=
	j^i.
\end{equation}
Using the expression \eqref{current} for the current, the Bohmian velocity field is therefore
\begin{equation}
	v^i(q,t)
	=
	G^{ij}\,\partial_j S(q,t).
\end{equation}
Thus, the Bohmian guidance equations are
\begin{equation}
		\dot q^i(t)
		=
		G^{ij}\,\partial_j S\big(q(t),t\big).
	\label{guidance}
\end{equation}
To assess the role of quantum effects, we compare these Bohmian trajectories with the corresponding classical trajectories obtained from
\begin{equation}
	\dot{{q}} = \frac{\partial H}{\partial {p}}, \qquad \dot{{p}} = - \frac{\partial H}{\partial {q}} 
	\label{clequm}
\end{equation}
with the same initial position ${q}(0)$ as the Bohmian particle and its momentum chosen as ${p}(0) = \nabla S({q}(0),0)$.

\subsection{Relation to higher time-derivative theories}

The connection with HTDTs is made through the standard reduction to first-order form. Starting from a higher-derivative Lagrangian
\[
L(q,\dot q,\ddot q,\ldots),
\]
one introduces auxiliary variables so that the dynamics is rewritten as an equivalent first-order system on an enlarged configuration space. In the simplest case of a second-order reduction one may take
\[
Q=(q,v), \qquad v=\dot q,
\]
and formulate the corresponding Schr\"odinger problem in these reduced variables. The resulting Hamiltonian typically contains a ghost sector because it is linear in at least one canonical momentum, thereby inheriting the characteristic Ostrogradsky-type unboundedness.

From the Bohmian viewpoint, however, the reduced system can be treated in the same structural way as the two-dimensional ghost Hamiltonian considered above. Writing the wavefunction on the enlarged configuration space in polar form as
\[
\Psi(Q,t)=R(Q,t)e^{iS(Q,t)/\hbar},
\]
the guidance law is again determined by the gradient of the phase \(S(Q,t)\), composed with the kinetic tensor appearing in the reduced Hamiltonian. Thus the Bohmian flow provides a direct dynamical probe of the reduced higher-derivative system, allowing one to compare packet transport, spreading, and instability in a way that is sensitive to the chosen Hamiltonian representation.

This perspective is especially useful in settings where different first-order reductions or Hamiltonian representations generate the same classical equations of motion. As we shall see below, such classical equivalence need not imply Bohmian quantum equivalence, since the guidance law retains information about the specific Hamiltonian structure used in the quantisation. The trajectory-based formulation therefore provides additional information beyond the classical phase-space dynamics alone, and is particularly well suited to the analysis of higher time-derivative models with nonstandard kinetic structure.

\section{Diagnostics of a coupled ghost oscillator model}

We consider the two-dimensional ghost Hamiltonian with Lorentzian kinetic term introduced in \cite{fring2025quant,FFT},
\begin{equation}
	H_{\mathrm{gh}}(x,y,p_x,p_y)
	=
	\frac{1}{2}\big(p_x^2-p_y^2\big)
	+\nu^2 x^2
	+\Omega y^2
	+g xy,
	\qquad \nu,\Omega,g\in\mathbb{R} ,
	\label{Hghost}
\end{equation}
with corresponding Schr\"odinger equation
\begin{equation}
	i\hbar\,\partial_t\psi(x,y,t)
	=
	\left[
	-\frac{\hbar^2}{2}\big(\partial_x^2-\partial_y^2\big)
	+V_{\mathrm{gh}}(x,y)
	\right]\psi(x,y,t),
	\label{SchroedingerGhost}
\end{equation}
where
\begin{equation}
	V_{\mathrm{gh}}(x,y)
	= \frac{1}{2} q^\intercal C q,
	\qquad
	C=
	\begin{pmatrix}
		2\nu^2 & g \\
		g & 2\Omega
	\end{pmatrix}.
	\label{Vghost}
\end{equation}
To probe nontrivial Bohmian dynamics, we work with Gaussian wavepackets rather than stationary eigenstates.

\subsection{Gaussian wavepackets and diagnostic quantities}
We begin with a Gaussian wavepacket of the form
\begin{equation}
	\psi_g(q,t)
	=
	\mathcal N(t)\,
	\exp\!\left[
	-\frac{1}{2\hbar}(q-q_c)^\intercal (A+iB)(q-q_c)
	+\frac{i}{\hbar}\,p_c\!\cdot\!(q-q_c)
	+\frac{i}{\hbar}\,\theta(t)
	\right],
	\label{Gaussianpacket}
\end{equation}
where \(A(t)\) and \(B(t)\) are real symmetric matrices, \(q_c(t)\) is the configuration-space centre, and \(p_c(t)\) the corresponding central momentum. Substituting \eqref{Gaussianpacket} into \eqref{SchroedingerGhost} yields
\begin{equation}
	\dot q_c = G p_c,
	\qquad
	\dot p_c = -C q_c,
	\label{centreequ}
\end{equation}
together with the Riccati equation
\begin{equation}
	\dot K = - i (K G K - C),
	\qquad K=A+iB,
	\label{Ricatti}
\end{equation}
together with evolution equations for \(\theta(t)\) and \(\mathcal N(t)\), which determine the overall phase and normalisation but do not affect the trajectory analysis directly. Since \eqref{centreequ} coincides with the classical equations of motion,
\begin{equation}
	\dot q = Gp,
	\qquad
	\dot p = -Cq,
	\label{classicaleqm}
\end{equation}
the centre of the Gaussian packet follows the classical trajectory exactly, as expected for quadratic Hamiltonians, see e.g. \cite{sakurai2020modern,klauder1985coherent}.

Factorising $\psi_g$ in (\ref{Gaussianpacket}) into the polar form (\ref{WKB}) using
\begin{equation}
	R \propto 
	\exp\!\left[-\frac{1}{2\hbar}(q-q_c)^{\intercal}A(q-q_c)\right], \quad
	S= 
	-\frac{1}{2}
	({q}-{q}_c)^{\intercal}
	B
	({q}-{q}_c)
	+
	{p}_c \cdot ({q}-{q}_c)
	+ \theta,
\end{equation}
we obtain with (\ref{Vghost}) the Bohmian guidance law
\begin{equation}
	\dot q = G\nabla S,
	\qquad
	\nabla S(q,t)=p_c-B(q-q_c),
	\label{Bohmtraj-section3}
\end{equation}
together with the quantum potential
\begin{equation}
	Q(q,t)
	=
	-\frac12 (q-q_c)^\intercal AGA(q-q_c)
	+\frac{\hbar}{2} \text{Tr}(GA).
	\label{exquantumpot}
\end{equation}
It is convenient to introduce the deviation from the packet centre,
\begin{equation}
u(t):=q_B(t)-q_c(t),
\end{equation}
and the quantum-classical separation,
\begin{equation}
\Delta(t):=q_B(t)-q_{\mathrm{cl}}(t).
\end{equation}
Using \eqref{Bohmtraj-section3} and \eqref{centreequ}, the deviation satisfies the first-order equation
\begin{equation}
	\dot u = -GB\,u.
	\label{intfirstord}
\end{equation}
Thus the internal Bohmian flow is controlled by the matrix
\begin{equation}
	M(t):=-GB(t),
	\label{Mmatrix}
\end{equation}
whose symmetric part
\begin{equation}
	S_M(t):=\frac12\bigl(M(t)+M^\intercal(t)\bigr)
	\label{SMmatrix}
\end{equation}
controls the instantaneous growth or decay of the deviation \(u(t)\). This is seen from
\begin{equation}
	\frac{d}{dt}\|u(t)\|^2
	= \frac{d}{dt} \left( u^\intercal u    \right) = \dot{u}^\intercal u+u^\intercal \dot{u}=
	u^\intercal   \left( M^\intercal + M   \right) u =
	2u^\intercal S_M(t)\,u ,
	\label{norm-growth}
\end{equation}
so that positive eigenvalues of \(S_M(t)\) correspond to locally expanding directions in the Bohmian flow, while negative eigenvalues correspond to contracting directions. The antisymmetric part of \(M(t)\) contributes only to the rotational part of the motion and does not directly change the magnitude of \(u(t)\).

Differentiating \eqref{intfirstord} and using \eqref{Ricatti} gives the equivalent second-order equation
\begin{equation}
	\ddot u = -G\Lambda\,u,
	\qquad
	\Lambda:=C-AGA.
	\label{intersecord}
\end{equation}
The matrix \(\Lambda\) measures the mismatch between the classical curvature \(C\) and the quantum curvature \(AGA\). When \(\Lambda=0\), the internal acceleration vanishes and the packet undergoes rigid transport and when \(\Lambda\neq0\), Bohmian trajectories deform relative to the centre.

These equations provide a compact classification criterion. Since the Bohmian trajectory can be written as $q_B(t)=q_c(t)+u(t)$, bounded motion requires control of both the packet centre and the internal deviation, as well as admissibility of the Gaussian state itself. First, the packet must remain normalisable, which in the present parametrisation means that the amplitude matrix $A$ remains positive definite. Second, the centre trajectory $q_c(t)$ must stay bounded, since otherwise the full Bohmian motion inherits the runaway of the packet centre. Third, the internal flow generated by $-GB$ must avoid sustained expanding directions, so that the deviation $u(t)$ does not grow secularly. Failure of these three conditions leads respectively to non-normalisability of the state, drift-induced runaway of the full trajectory, or internally amplified Bohmian instability.

A useful structural observation is that exact cancellation, \(\Lambda=0\), can occur only for non-confining saddle-type potentials. Indeed, \(\Lambda=0\) implies \(C=AGA\), and hence
\begin{equation}
	\det(AGA)=\det(G)\det(A)^2=-\det(A)^2<0,
\end{equation}
so \(AGA\), and therefore \(C\), must be indefinite:
\begin{equation}
	\det C = 4\nu^2\Omega-g^2<0.
\end{equation}
Thus, if $C$ is positive definite, one necessarily has \(\Lambda\neq0\), so no Gaussian state can exactly cancel the classical curvature. In this sense, the mixed signature of \(G\) obstructs the usual coherent-state cancellation mechanism familiar from positive-definite systems.

In the numerical examples below we compare three families of trajectories: the classical ensemble obtained from \eqref{classicaleqm}, the Gaussian centre determined by \eqref{centreequ}, and the Bohmian ensemble obtained from \eqref{Bohmtraj-section3}. For the ensemble plots, the initial positions are sampled from the initial density \(|\psi(q,0)|^2\), while the classical initial momenta are chosen from the initial phase gradient \(p(0)=\nabla S(q,0)\). 

In the discussion below, boundedness or growth of \(u(t)\) is interpreted through the internal flow \(\dot u=-GB\,u\), while the behaviour of \(\Lambda(t)\) indicates whether the Bohmian dynamics remains close to the rigid-transport limit or develops significant curvature-driven deformation.

\subsubsection{Rigid-transport regime}

\begin{figure}[h]
	\begin{minipage}[b]{\textwidth}      
		\centering
		\includegraphics[width=0.48\textwidth]{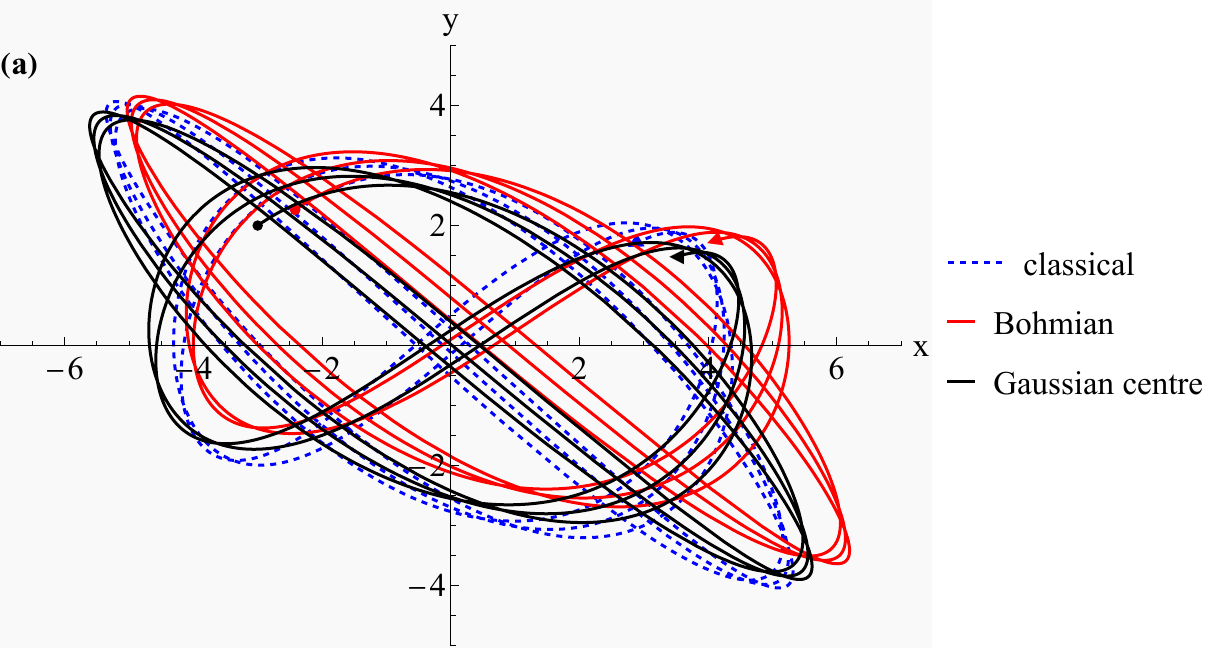}
		\includegraphics[width=0.42\textwidth]{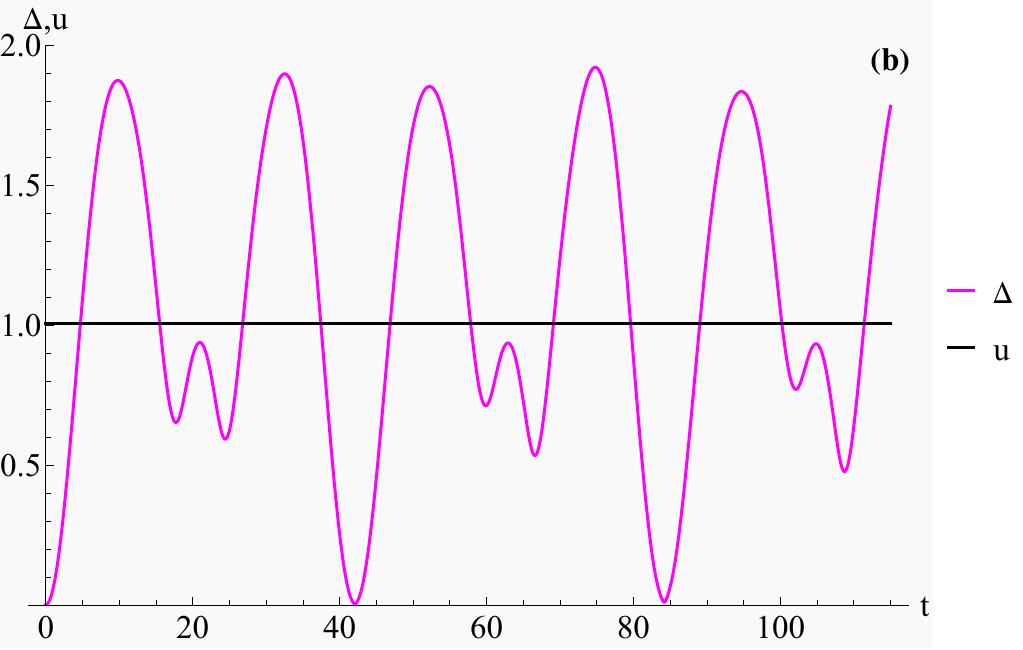}
	\end{minipage}   
	\caption{
		Phase-space trajectories and quantum-classical diagnostics in the rigid-transport regime.
		(a) Phase-space projections of three trajectory families: classical ensemble (blue dashed), Bohmian trajectories (red solid), and the Gaussian packet centre (black solid). The initial Gaussian wavepacket parameters are $A_{11}(0)=1/2\sigma_x^2$, $A_{22}(0)=1/2\sigma_y^2$, $A_{12}(0)=A_{21}(0)=0.2$, with $\sigma_x=1.2$, $\sigma_y=1.0$, $q_c(0)=(-3,2)$, $p_c(0)=(1,-0.75)$, and $B(0)=0$. The initial and final points, at $t=0$ and $t=115$, are marked by a filled circle and a triangle, respectively. (b) Time series of the internal deviation $u(t)=q_{\mathrm{Bohm}}(t)-q_c(t)$ (black) and the quantum–classical separation $\Delta(t)=q_{\mathrm{Bohm}}(t)-q_{\mathrm{classical}}(t)$ (magenta), averaged over the Bohmian ensemble. The potential is characterised by $\nu =0.200703$, $\Omega = -0.105$ and $g=-0.0305556$.}
	\label{rigid}
\end{figure}

We begin with a regime in which the curvature mismatch \(\Lambda(t)\) vanishes exactly or remains sufficiently small over the time interval considered that the internal acceleration in \eqref{intersecord} is negligible. The results for this regime are shown in figure \ref{rigid}.

Panel (a) shows that the classical ensemble, the Bohmian ensemble, and the Gaussian centre all remain bounded and follow closely related loop-like trajectories. Panel (b) shows that the internal deviation \(u(t)\) remains approximately constant, while the quantum-classical separation \(\Delta(t)\) exhibits only bounded oscillations. As explained in section 3.1, this indicates that the internal flow generated by \(-GB\) does not develop sustained expanding directions, and that the curvature mismatch is too weak to produce secular growth through \eqref{intersecord}. The packet is therefore transported with little internal deformation, so this regime is naturally interpreted as coherent or near-coherent rigid transport.

\subsubsection{The quasi-semiclassical regime $\Lambda \neq 0$}
\begin{figure}[h]
	\begin{minipage}[b]{0.66\textwidth}
		\centering
		\includegraphics[width=\textwidth]{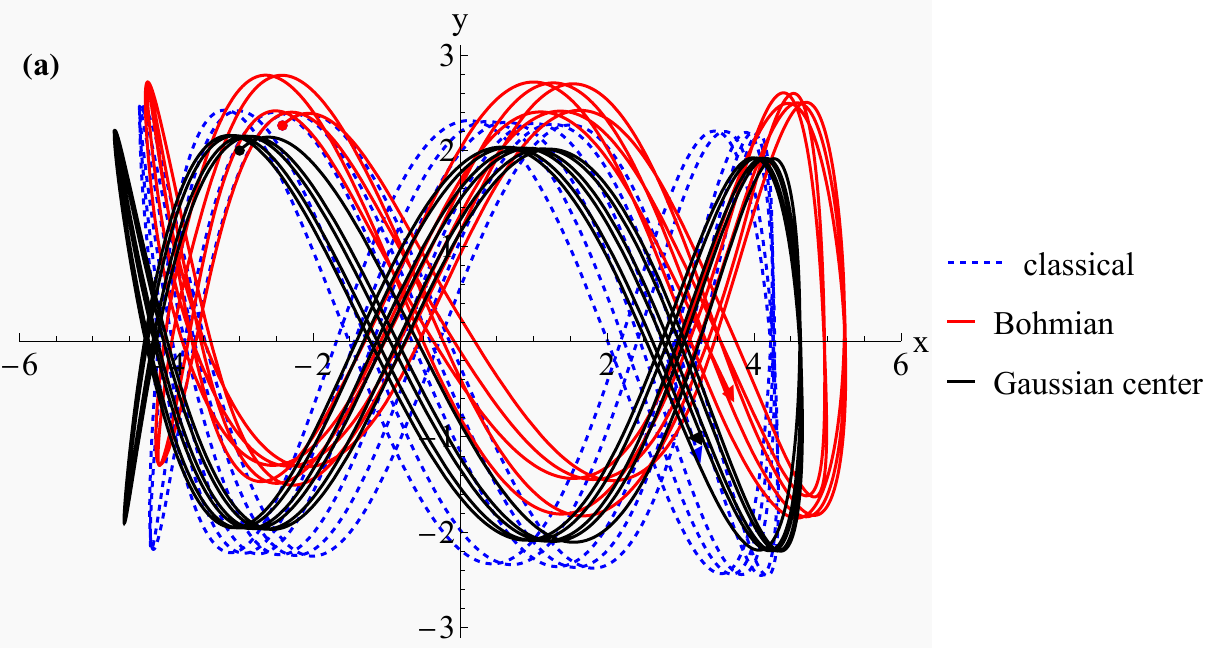}
		\vspace{1.em}
	\end{minipage}
	\hfill
	\begin{minipage}[b]{0.32\textwidth}
		\centering
		\includegraphics[width=\textwidth]{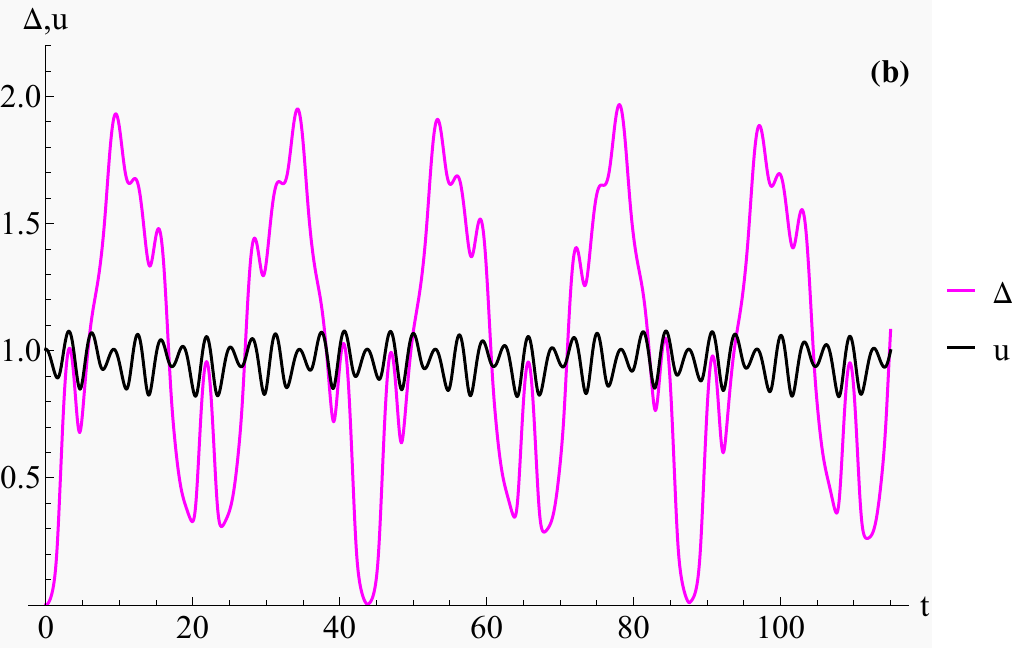}
		\vspace{0.5em}
		\includegraphics[width=\textwidth]{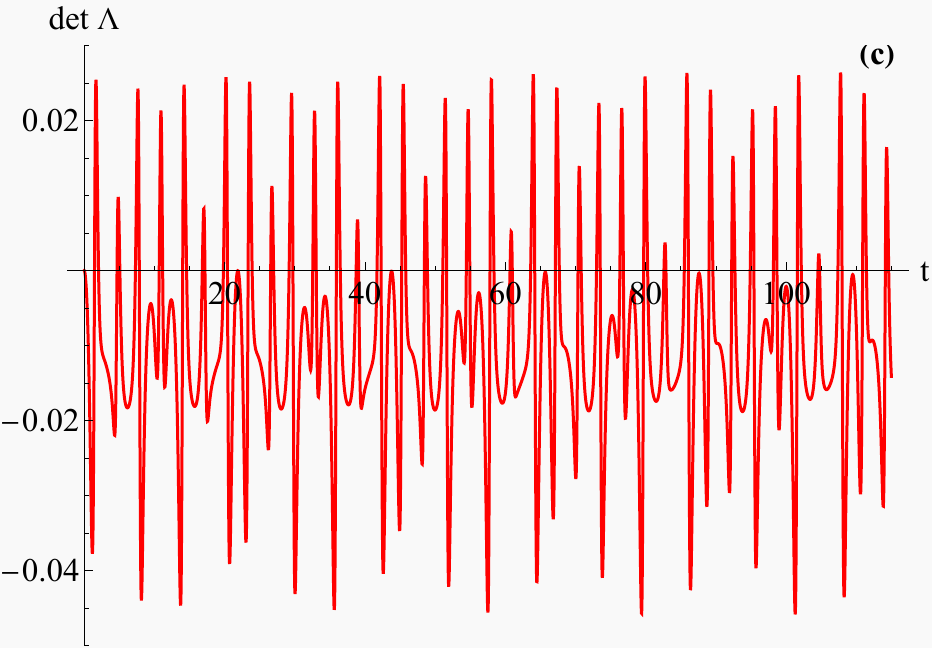}
	\end{minipage}
	\caption{Phase-space trajectories and quantum-classical diagnostics in the quasi-semiclassical regime (a) for the classical ensemble (blue dashed), Bohmian ensemble (red), and Gaussian centre (black) in a regime with $\Omega$ reduced by $0.4$ relative to figure \ref{rigid}. (b) Time series of the internal deviation $u(t)$ (black) and quantum–classical separation $\Delta(t)$ (magenta).  (c) Time evolution of $\det\Lambda(t)$.}
	\label{quasisemicl}
\end{figure}

We next consider a regime in which \(\Lambda(t)\neq0\), but remains bounded and oscillatory. The corresponding trajectories are shown in figure \ref{quasisemicl}, obtained by varying only the frequency parameter \(\Omega\) relative to the rigid-transport case.

Panel (a) shows that the classical ensemble, the Gaussian centre, and the Bohmian ensemble all remain bounded. Panel (b) confirms that neither the internal deviation \(u(t)\) nor the quantum-classical separation \(\Delta(t)\) develops secular growth. Panel (c) shows that \(\det\Lambda(t)\) fluctuates about zero rather than remaining close to zero, so the packet is no longer in the strict rigid-transport limit. Nevertheless, these oscillations do not translate into persistent expanding directions in the internal Bohmian flow, and the resulting motion remains bounded despite the non-vanishing curvature mismatch. This regime is therefore no longer coherent in the strict \(\Lambda=0\) sense, but remains quasi-semiclassical.

In this sense, the present Bohmian analysis extends the classical observations of \cite{deffayet2022ghosts,deffayet2023global,diez2024foundations} to the quantum domain. Those works showed that ghost degrees of freedom need not lead to runaway instabilities at the classical level when the coupled system is suitably structured. Here we find an analogous phenomenon for Gaussian Bohmian dynamics: despite the indefinite kinetic structure, both the packet centre and the Bohmian ensemble can remain bounded, with the bounded regime characterised by controlled curvature mismatch and the absence of persistent expansion in the internal flow.

\begin{figure}[h]
	\begin{minipage}[b]{0.66\textwidth}
		\centering
		\includegraphics[width=\textwidth]{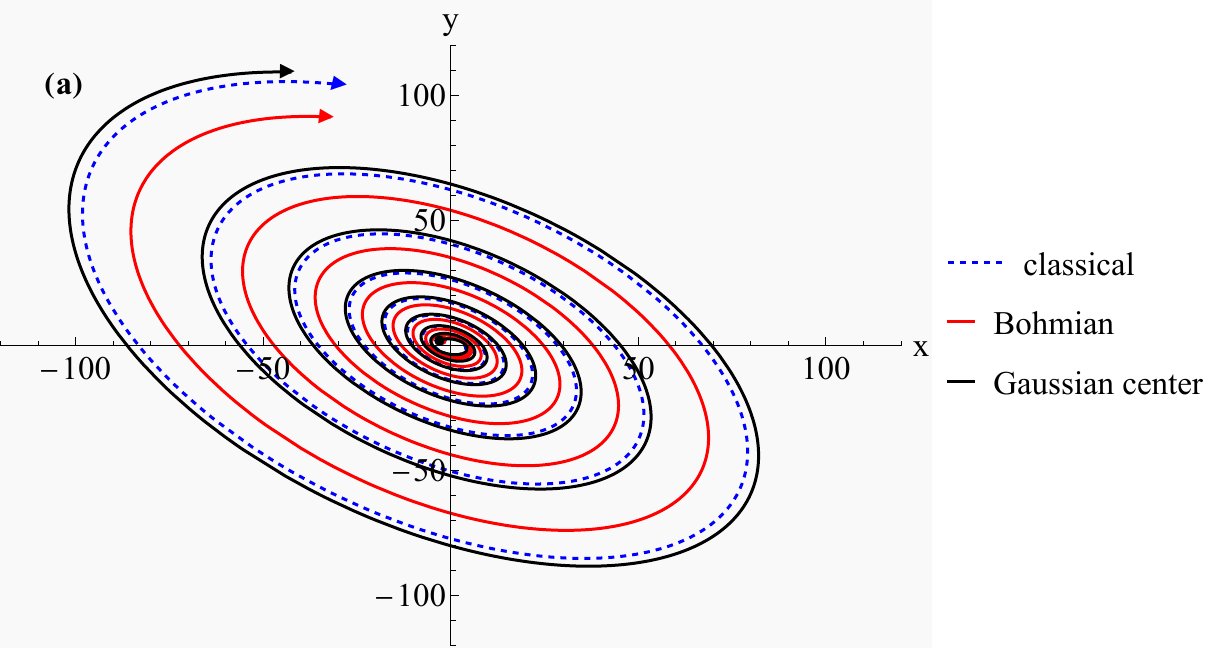}
		\vspace{1.em}
	\end{minipage}
	\hfill
	\begin{minipage}[b]{0.32\textwidth}
		\centering
		\includegraphics[width=\textwidth]{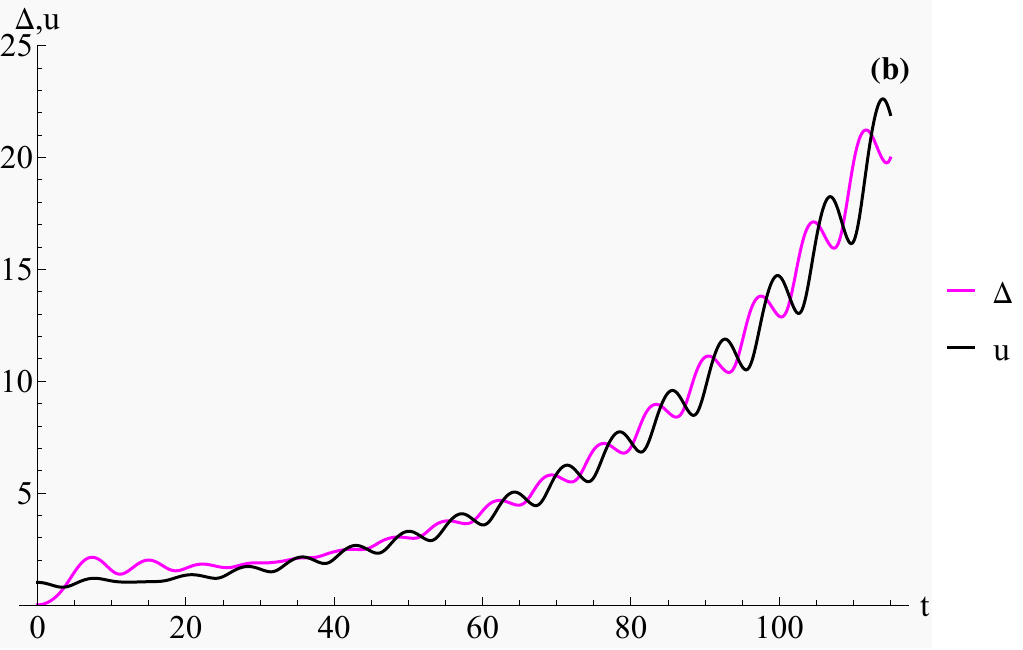}
		\vspace{0.5em}
		\includegraphics[width=\textwidth]{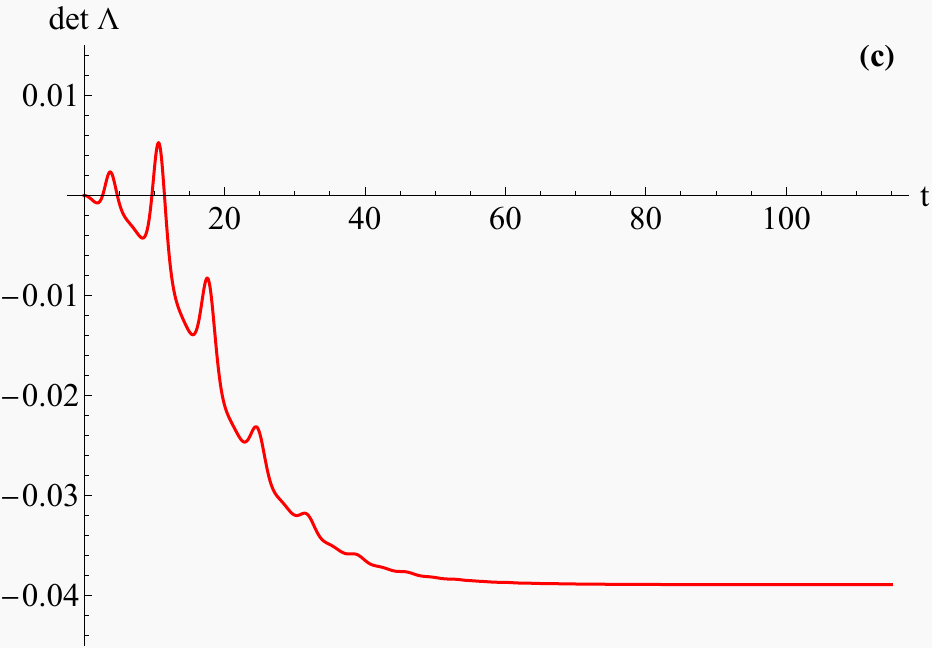}
	\end{minipage}
	\caption{Phase-space trajectories in the unstable spiral regime. Panel (a) shows the classical ensemble (blue dashed), Bohmian ensemble (red), and Gaussian centre (black) in a regime with $\nu$ reduced by $0.1$ relative to figure \ref{rigid}. (b) Time series of the internal deviation $u(t)$ (green) and quantum–classical separation $\Delta(t)$ (magenta).  (c) Time evolution of $\det\Lambda(t)$.}
	\label{figure5}
\end{figure}

\subsubsection{The unstable spiral regime} 
In the unstable spiral regime, the linearised centre dynamics possesses complex-conjugate eigenvalues with positive real part, so the centre motion is already unbounded. The resulting trajectories are shown in figure \ref{figure5}.

Panel (a) shows that the classical ensemble, the Gaussian centre, and the Bohmian ensemble all spiral outward with increasing amplitude. Panel (b) indicates that both the internal deviation \(u(t)\) and the quantum-classical separation \(\Delta(t)\) grow in time, so the Bohmian flow not only follows the unstable centre motion but also develops increasing separation relative to both the packet centre and the corresponding classical trajectories. Panel (c), through the nontrivial evolution of \(\det\Lambda(t)\), shows that the curvature mismatch remains dynamically active throughout the spiralling phase. In terms of the diagnostics introduced in section 3.1, this regime combines unstable centre motion with expanding behaviour in the internal flow, leading to runaway dynamics with a rotational component.

\subsubsection{The critical regime}
\begin{figure}[h]
	\begin{minipage}[b]{0.66\textwidth}
		\centering
		\includegraphics[width=\textwidth]{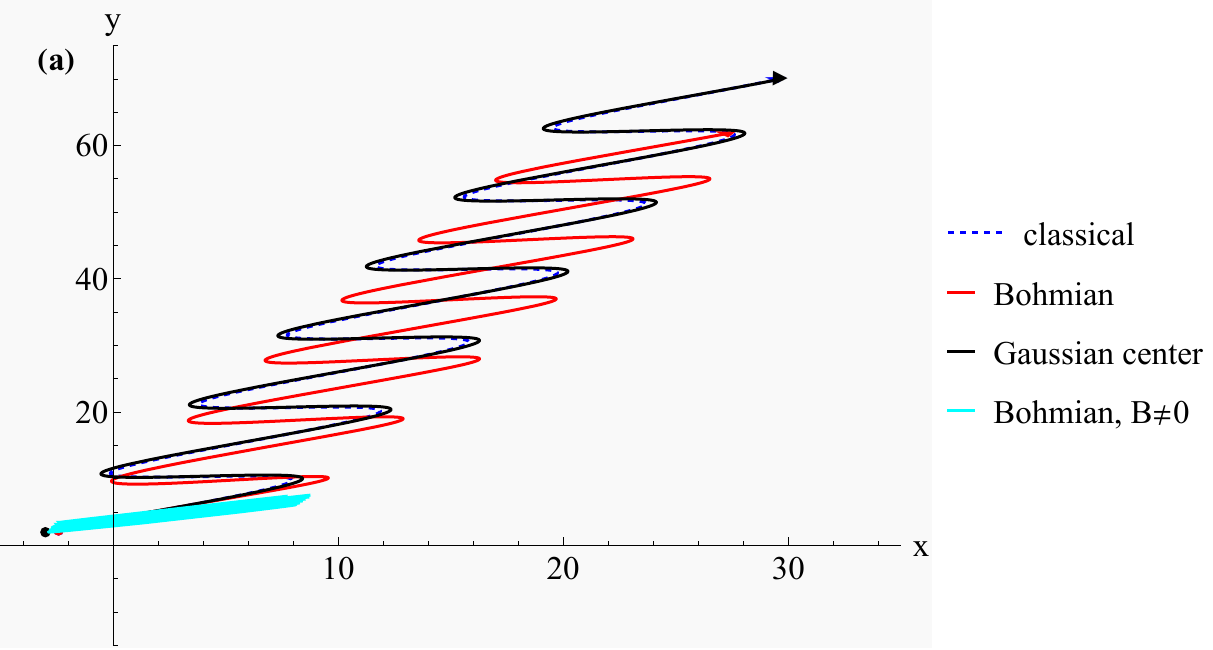}
		\vspace{1.em}
	\end{minipage}
	\hfill
	\begin{minipage}[b]{0.32\textwidth}
		\centering
		\includegraphics[width=\textwidth]{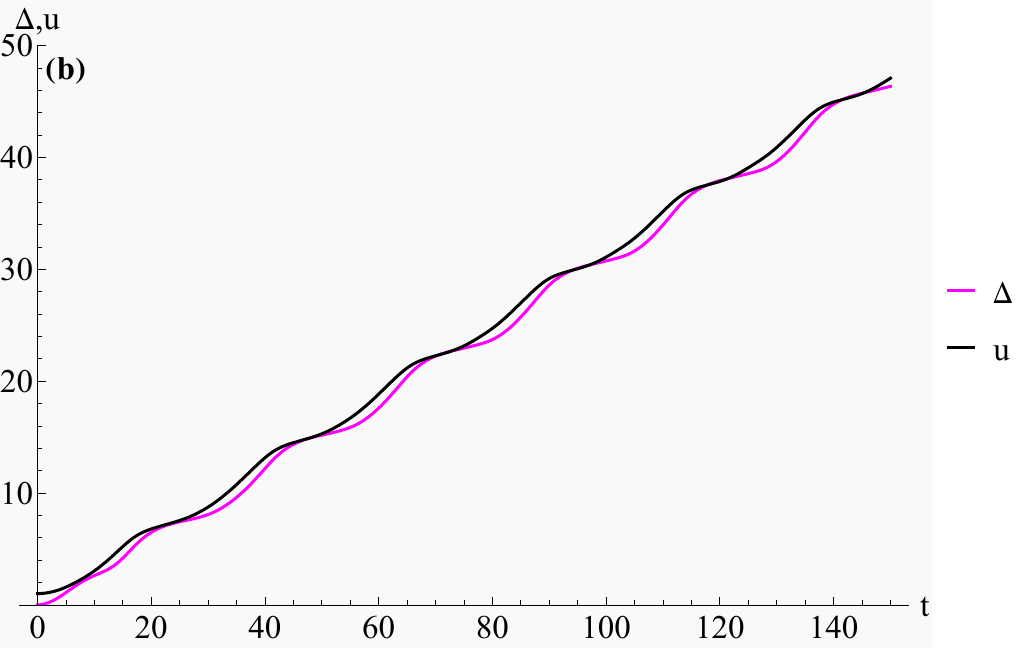}
		\vspace{0.5em}
		\includegraphics[width=\textwidth]{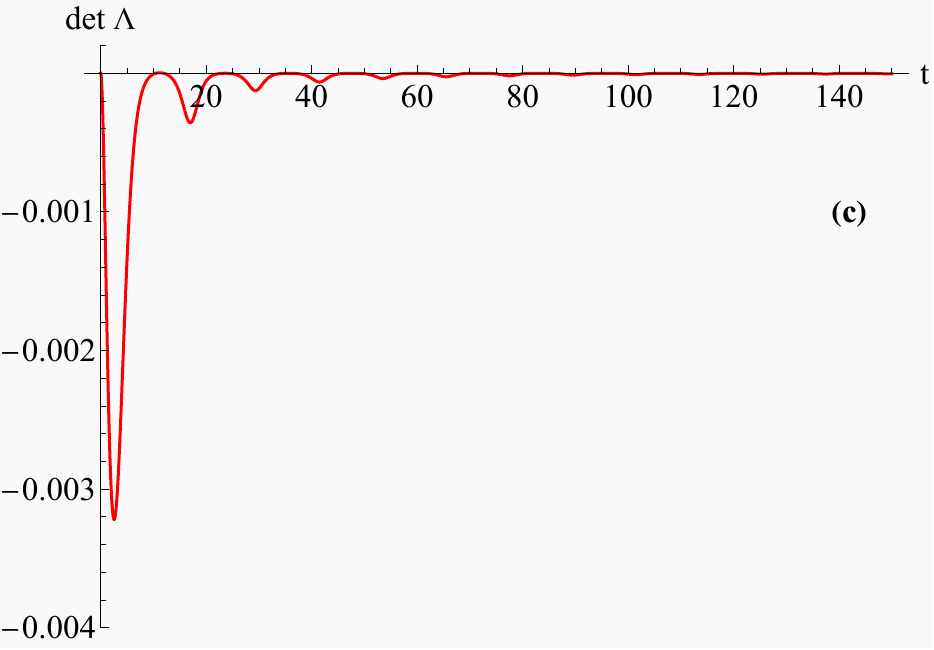}
	\end{minipage}
	\caption{Phase-space trajectories and quantum-classical diagnostics at the critical point $\det C =0$ (a) for the classical ensemble (blue dashed), Bohmian ensemble with $B=0$ (red), $B_{xx}=B_{yy}=0$, $B_{xy}=B_{yx}=-0.45$ (cyan), and Gaussian centre (black). (b) Time series of the internal deviation $u(t)$ (black) and quantum–classical separation $\Delta(t)$ (magenta).  (c) Time evolution of $\det\Lambda(t)$. The potential is characterised by $\nu =0.200703$, $\Omega =0.00579446$ and $g=-0.0305556$. }
	\label{critpoint}
\end{figure}

We next explore the critical regime in which the potential curvature matrix becomes degenerate, \(\det C=0\), while the internal curvature remains nonzero. At this point the system loses one restoring direction, so the Gaussian centre is no longer confined. The resulting trajectories are shown in figure \ref{critpoint}.

Panel (a) shows that all three trajectory families become unbounded. Panel (b) demonstrates that both the internal deviation \(u(t)\) and the quantum-classical separation \(\Delta(t)\) now grow rather than oscillate, signalling the breakdown of bounded transport. Panel (c) shows that \(\det\Lambda(t)\) remains nonzero, so the curvature mismatch persists throughout the evolution and the dynamics does not approach the rigid-transport limit. The Bohmian trajectories typically diverge somewhat more slowly than the classical ones, especially when an initial chirp is included, but this only moderates the instability rather than removing it. According to the criteria established in section 3.1, the critical case therefore marks the transition from bounded motion to runaway behaviour.

\subsubsection{Non-normalisable initial wavepacket}
In models with ghost sectors of the type studied here, quantisation naturally produces multiple sectors with distinct spectral and localisation properties. As discussed in \cite{fring2025quant,FFT}, some sectors exhibit spectra bounded from below but correspond to eigenfunctions that are not square-integrable. While such non-normalisable states are typically excluded in standard quantum mechanics because they do not belong to the Hilbert space \(L^2(\mathbb{R}^n)\), they can nevertheless be useful as formal test functions or asymptotic approximations, e.g. plane waves in scattering theory, and still generate well-defined Bohmian velocity fields \cite{sen2024physical}. Bohmian mechanics only requires a differentiable complex wavefunction to define a velocity field according to (\ref{guidance}) without requiring a strictly normalisable probability density \cite{Bohm1952,Holland1993}.

\begin{figure}[h]
	\begin{minipage}[b]{\textwidth}      
		\centering
		\includegraphics[width=0.48\textwidth]{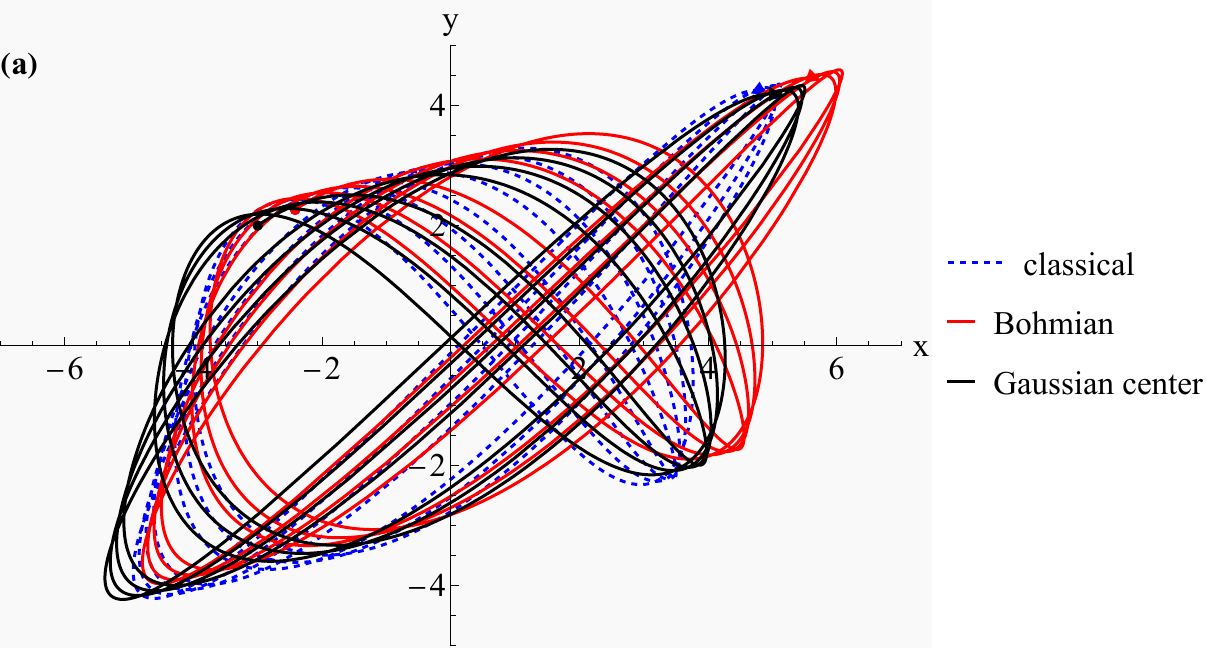}
		\includegraphics[width=0.42\textwidth]{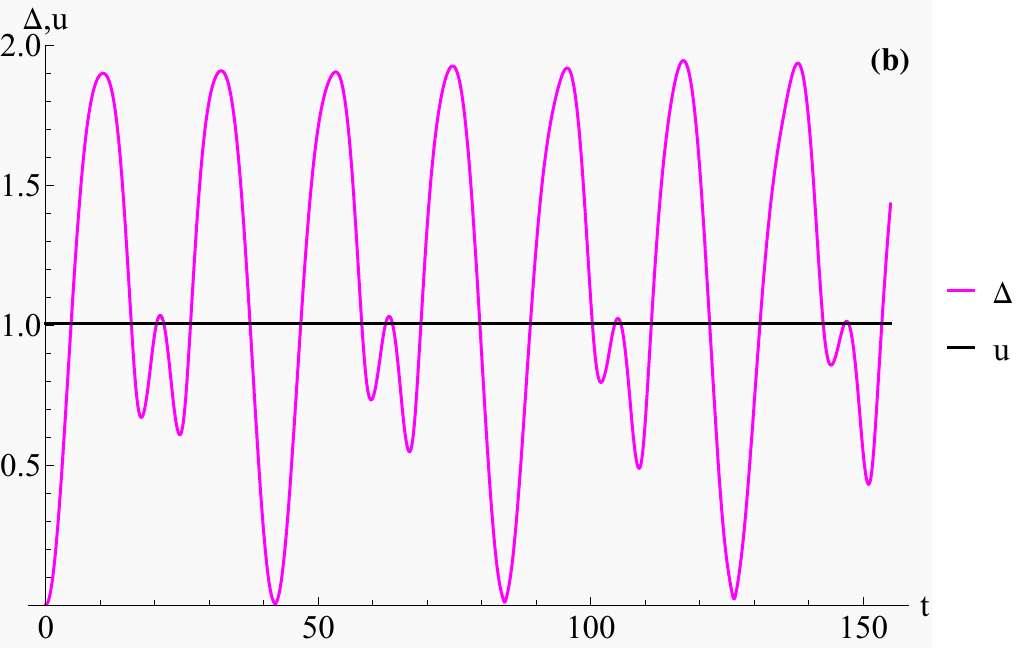}
	\end{minipage}   
	\caption{
		Phase-space trajectories and diagnostics in the rigid-transport regime for a non-normalisable Gaussian wavepacket. (a) Classical ensemble (blue dashed), Bohmian ensemble (red), and Gaussian centre (black). (b) Time series of internal deviation \(u(t)\) (green) and quantum–classical separation \(\Delta(t)\) (magenta). The initial Gaussian wavepacket parameters are $A_{11}(0)=-1/2\sigma_x^2$, $A_{22}(0)=-1/2\sigma_y^2$, $A_{12}(0)=A_{21}(0)=c$ with the remaining ones identical to those in figure  \ref{rigid}. The potential is characterised by $\nu =0.200703$, $\Omega = -0.105$ and $g=0.0305556$.
	}
	\label{nonnormrigid}
\end{figure}

To investigate this possibility, we choose the initial spread matrix \(A(0)\) so that its real symmetric part has indefinite sign, yielding a non-normalisable Gaussian packet, while retaining the rigid-transport conditions on the potential. The resulting trajectories are shown in figure \ref{nonnormrigid}.

Panel (a) shows that the classical ensemble, the Bohmian ensemble, and the Gaussian centre continue to trace bounded orbits with qualitative features similar to those of the normalisable rigid-transport regime. Panel (b) shows that the internal deviation \(u(t)\) remains approximately constant while \(\Delta(t)\) oscillates in a bounded manner. In the terminology of section 3.1, the internal flow therefore remains non-expanding, even though the state fails the normalisability condition required for a physical Gaussian packet. This suggests that the phase-curvature cancellation mechanism underlying rigid transport is largely insensitive to square-integrability, although the resulting state should be regarded only as a diagnostic probe rather than a physical \(L^2\) state.

The different cases studied above are summarised in table \ref{tab:regimes}, which organises the numerical results according to normalisability, centre dynamics, and internal Bohmian stability.

\begin{table}[h]
	\centering
	\small
	\setlength{\tabcolsep}{4pt}
	\begin{tabular}{|p{2.6cm}|p{3.0cm}|p{3.2cm}|p{3.4cm}|}
		\hline
		Regime & Conditions & Bohmian behaviour & Interpretation \\
		\hline \hline
		Rigid transport &
		$A(t)>0$, $\Lambda(t) \approx 0$ or $\|\Lambda(t)\|\ll1$ &
		$q_c(t)$ bounded; $u(t)$ and $\Delta(t)$ remain bounded &
		coherent or near-coherent transport \\
		\hline
		Quasi-semiclassical &
		$A(t)>0$, $\Lambda(t)\neq0$ but bounded &
		bounded centre motion with oscillatory internal deformation &
		breathing/shearing packet without secular growth \\
		\hline
		Spiral instability &
		$A(t)>0$ initially; expanding directions in $-G B(t)$ &
		rotational flow with growing $u(t)$ and $\Delta(t)$ &
		spiral instability with quantum deformation \\
		\hline
		Critical runaway &
		$\det C=0$, $\Lambda(t)\neq0$ &
		marginal behaviour followed by linear or accelerated growth &
		transition to runaway motion \\
		\hline
		Non-normalisable sector &
		$A(t)$ loses positive definiteness &
		formal Bohmian flow may persist &
		diagnostic only, not a physical $L^2$ state \\
		\hline
	\end{tabular}
	\caption{Summary of the dynamical regimes studied in section 3. Here \(A(t)\) controls normalisability, \(B(t)\) the local Bohmian flow, and \(\Lambda(t)=C-AGA\) the curvature mismatch.}
	\label{tab:regimes}
\end{table}

\section{Quantum ambiguities}   

As discussed in the literature on the PU model, the ghost Hamiltonian formulation admits a bi-Hamiltonian partner that generates the same classical flow, while leading to a different quantum realisation, see in particular \cite{FFT}, and also the discussion of the hidden symmetry structure at the degenerate point in \cite{fring2026spec}. In this section we use the Bohmian framework to analyse this ambiguity dynamically.

Following \cite{fring2026spec}, dropping for convenience the overall factor \(1/2\) in the ghost Hamiltonian, we consider the pair
\begin{equation}
	H_g(x,y,p_x,p_y)
	=
	p_x^2-p_y^2+\nu^2x^2+\Omega y^2-(\nu^2+\Omega)xy,
	\label{Hgbiham}
\end{equation}
and
\begin{eqnarray}
	H_2 &=& \frac{1}{2 \sqrt{2}} \left[ \frac{\nu ^2-3 \Omega  }{\nu ^2-\Omega } p_x^2 - 2 \frac{ \nu ^2+\Omega  }{\nu ^2-\Omega } p_x p_y +\frac{\Omega -3 \nu ^2
	}{\nu ^2-\Omega }  p_y^2   +\frac{1}{2}  \left(3 \nu ^2-\Omega  \right)x^2 	-\left(\nu ^2+\Omega \right) x y\right.     \qquad \\ &&
	\left.   \qquad \qquad \qquad \qquad \qquad \qquad \qquad \qquad  \qquad \qquad \qquad \qquad \qquad \qquad+\frac{1}{2}  \left(3 \Omega -\nu ^2\right) y^2  \right] \notag 
\end{eqnarray}
together with the Poisson tensors
\begin{equation}
	J_g= \left(
	\begin{array}{cccc}
		0 & 0 & 1 & 0 \\
		0 & 0 & 0 & 1 \\
		-1 & 0 & 0 & 0 \\
		0 & -1 & 0 & 0 
	\end{array}
	\right) ,  \quad
	J_2= \frac{ 1  }{\sqrt{2} \left(\nu ^2-\Omega \right)}  \left(
	\begin{array}{cccc}
		0 & 0 & 3 \nu ^2-\Omega  & -\nu ^2-\Omega  \\
		0 & 0 & \nu ^2+\Omega  & \nu ^2-3 \Omega  \\
		\Omega -3 \nu ^2 & -\nu ^2-\Omega  & 0 & 0 \\
		\nu ^2+\Omega  & 3 \Omega -\nu ^2 & 0 & 0 
	\end{array}
	\right).
\end{equation}
These two Hamiltonian descriptions are classically equivalent in the sense that they generate the same phase-space vector field
\begin{equation}
	\dot z = J_g \nabla H_g = J_2 \nabla H_2,
	\qquad
	z=(x,y,p_x,p_y).
	\label{classequiv}
\end{equation}
Thus any difference found below is not classical in origin, but arises from the representation dependence of the corresponding Bohmian quantum theory.

This becomes explicit once the Gaussian ansatz \eqref{Gaussianpacket} is inserted into the guidance law. Although the two Hamiltonians generate the same classical equations of motion, they do so with different kinetic tensors, and these enter directly into the Bohmian velocity field. Writing the momentum-sector quadratic form as
\begin{equation}
	H_\alpha^{\mathrm{kin}}= p^\intercal G_\alpha p,
	\qquad
	\alpha\in\{g,2\},
\end{equation}
the corresponding tensors are
\begin{equation}
	G_g=
	\begin{pmatrix}
		1&0\\
		0&-1
	\end{pmatrix},
	\qquad
	G_2=
	\frac{1}{2 \sqrt{2}(\nu^2-\Omega)}
	\begin{pmatrix}
		\nu^2-3\Omega & -(\nu^2+\Omega)\\
		-(\nu^2+\Omega) & \Omega-3\nu^2
	\end{pmatrix}.
	\label{metrics-biham}
\end{equation}
Hence the Bohmian trajectories are obtained from
\begin{equation}
	\dot q_{B,\alpha}(t)=G_\alpha \nabla S_\alpha(q,t),
	\qquad
	\alpha\in\{g,2\},
	\label{guidance-biham}
\end{equation}
with the same initial positions sampled from the same initial packet, but with the guidance law determined in each case by the corresponding kinetic tensor \(G_\alpha\). Even when the initial phase profile is chosen identically, the subsequent Bohmian evolution differs because the velocity field depends explicitly on \(G_\alpha\). This is the precise origin of the representation dependence studied in this section.

\begin{figure}[h]
	\begin{minipage}[b]{\textwidth}      
		\begin{minipage}[b]{0.66\textwidth}
		\centering
		\includegraphics[width=\textwidth]{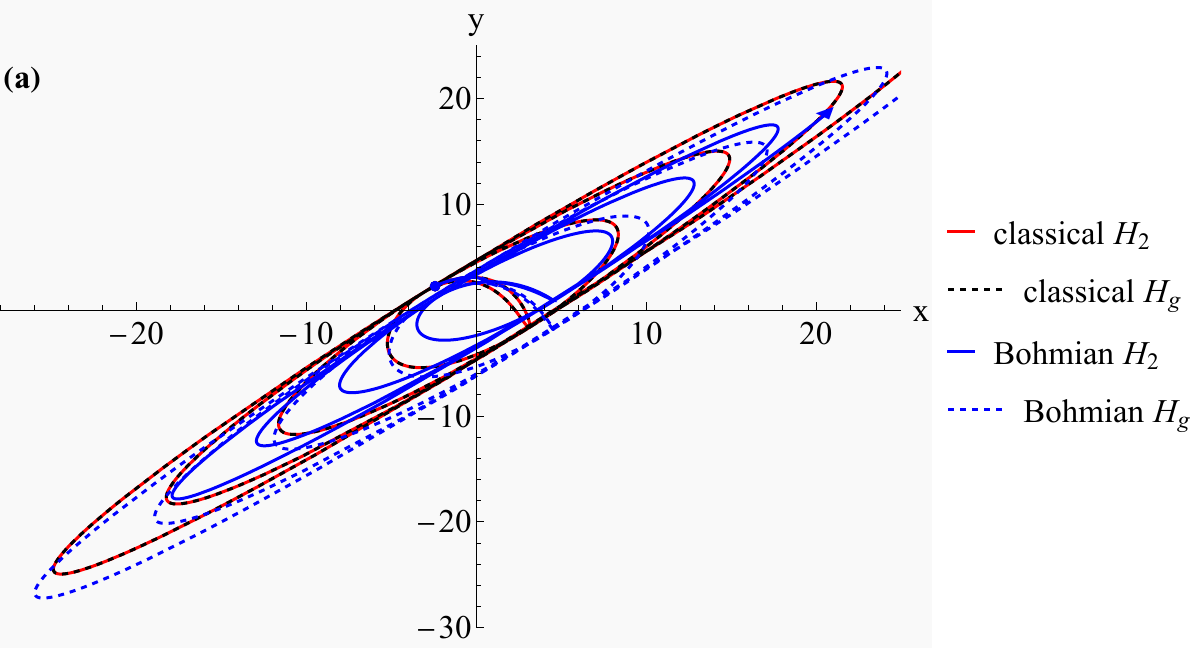}
		\vspace{1.em}
	\end{minipage}
	\hfill
	\begin{minipage}[b]{0.32\textwidth}
		\centering
		\includegraphics[width=\textwidth]{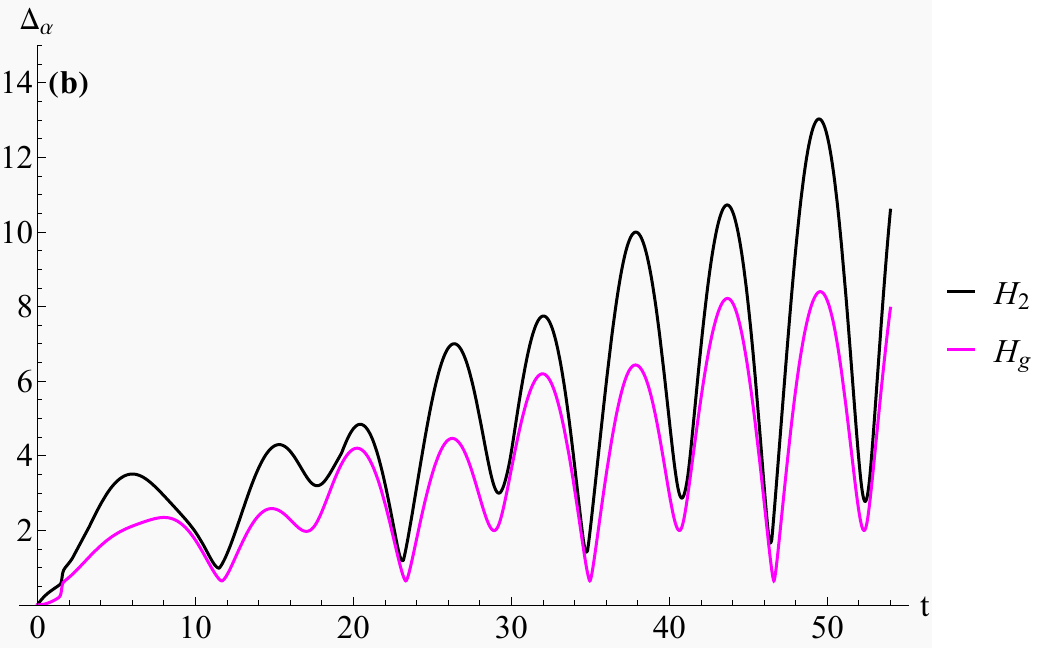}
		\vspace{0.5em}
		\includegraphics[width=\textwidth]{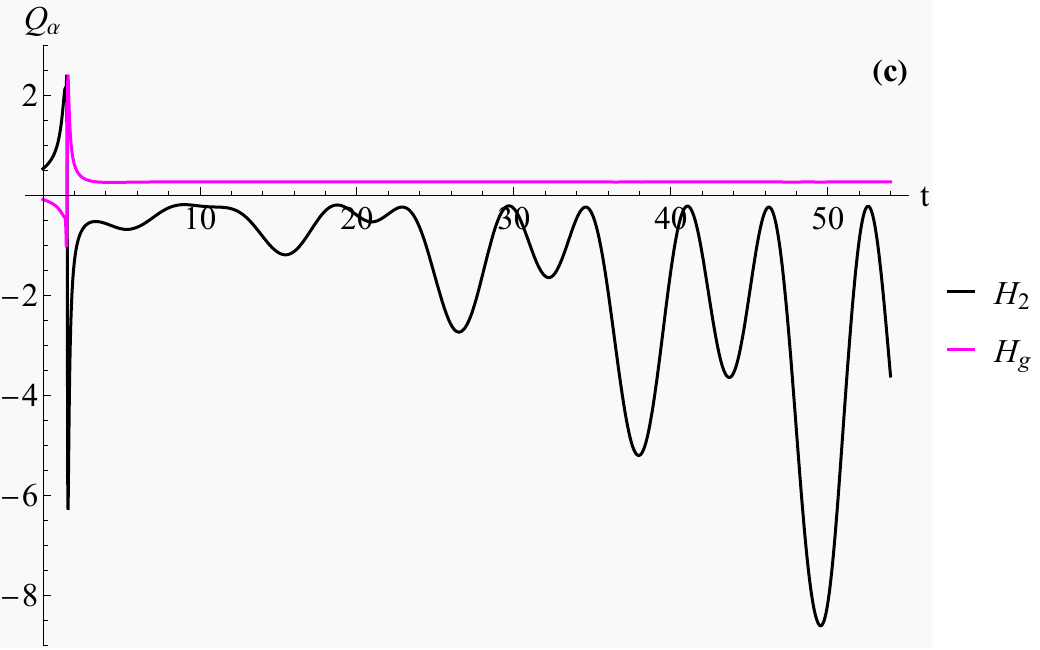}
	\end{minipage}
	\end{minipage}   
	\caption{Classical and Bohmian trajectories for the bi-Hamiltonian pair $H_g$ and $H_2$ in the degenerate regime for a non-normalisable Gaussian wavepacket. (a) Classical ensemble (black dashed) $H_g$, (red) $H_2$ and Bohmian ensemble (blue, dashed) $H_g$, (blue), $H_2$. (b) Time series of quantum–classical separation \(\Delta(t)_\alpha\), (c) quantum potential along the Bohmian trajectories \(Q(t)_\alpha\) for  $H_g$ (magenta) and $H_2$ (green). The initial Gaussian wavepacket parameters are $A_{11}(0)$, $A_{22}(0)$ as in figure \ref{rigid}, $A_{12}(0)=A_{21}(0)=0.2$, $B_{11}(0)=B_{22}(0)=0$, $B_{12}(0)=B_{21}(0)=0.01$. The model parameters are $\nu =0.200703$ and $\Omega = -0.105$. }
	\label{degpoint}
\end{figure}

The resulting phase-space trajectories are displayed in figure \ref{degpoint}(a). The classical ensembles generated by \(H_g\) and \(H_2\) coincide, in agreement with \eqref{classequiv}. By contrast, the Bohmian ensembles do not coincide. In both formulations the motion exhibits the same outward-spiralling runaway behaviour, as expected at the degenerate point of the PU oscillator, but the detailed Bohmian flow is visibly representation dependent. Thus the common classical instability is dressed by different quantum corrections in the two Hamiltonian descriptions.

A convenient way to quantify this effect is through the quantum-classical separation
\begin{equation}
	\Delta_\alpha(t):=
	q_{B,\alpha}(t)-q_{\mathrm{cl}}(t),
	\qquad
	\alpha\in\{g,2\},
	\label{Delta-alpha}
\end{equation}
where \(q_{\mathrm{cl}}(t)\) denotes the common classical trajectory with the same initial position and initial momentum fixed from the initial phase gradient. Figure \ref{degpoint}(b) shows the corresponding norms. In both cases the separation grows along the runaway motion, so the quantum correction does not remain a bounded perturbation of the classical orbit. Moreover, the growth is systematically stronger for \(H_2\) than for \(H_g\), showing that the second Hamiltonian representation induces a larger quantum departure from the same classical background.

Further insight is obtained from the quantum potential. For the Gaussian wavepacket considered here, the expression derived in (\ref{exquantumpot}) reads
\begin{equation}
	Q_\alpha(q,t)=
	-\frac12\bigl(q-q_{c,\alpha}(t)\bigr)^\intercal A_\alpha(t)G_\alpha A_\alpha(t)\bigl(q-q_{c,\alpha}(t)\bigr)
	+\frac{\hbar}{2} \text{Tr}\!\bigl(G_\alpha A_\alpha(t)\bigr),
	\quad
	\alpha\in\{g,2\}.
	\label{Qalpha}
\end{equation}
It is important to distinguish the field \(Q_\alpha(q,t)\) on configuration space from the quantity actually shown in figure \ref{degpoint}(c), namely the quantum potential evaluated along the Bohmian trajectory
\begin{equation}
	Q_{B,\alpha}(t):=
	Q_\alpha\!\bigl(q_{B,\alpha}(t),t\bigr).
	\label{QB-alpha}
\end{equation}
Because \(Q_\alpha\) depends explicitly on the kinetic tensor \(G_\alpha\), the two representations need not yield the same quantum potential even though they share the same classical flow. This is exactly what is seen in figure \ref{degpoint}(c). In the \(H_g\) case, \(Q_{B,g}(t)\) exhibits only a short initial transient before settling to an almost constant value. By contrast, in the \(H_2\) case the quantum potential remains strongly time-dependent and develops increasingly large oscillatory excursions. The two formulations therefore share the same classical runaway background, but differ significantly in the effective quantum force acting along the Bohmian trajectories.

Taken together, panels (a)-(c) of figure \ref{degpoint} provide a clear dynamical manifestation of the quantisation ambiguity already identified algebraically in \cite{FFT,fring2026spec}. The pair \((H_g,J_g)\) and \((H_2,J_2)\) is classically equivalent, but the associated Bohmian theories are not. The ambiguity is therefore genuinely quantum. It is invisible at the level of the classical phase-space vector field, but becomes manifest once one compares the Bohmian trajectories, the quantum-classical separation, and the quantum potential evaluated along the trajectories.

\section{Conclusions}

We analysed the Bohmian dynamics of a two-dimensional ghost Hamiltonian and its relation to HTDTs, with particular emphasis on Gaussian states and on the bi-Hamiltonian structure of the degenerate PU model. Our aim was not only to describe the corresponding quantum trajectories, but to test whether a trajectory-based formulation can reveal features that are less transparent at the level of spectral data or classical equations of motion alone.

For Gaussian wavepackets, the Bohmian formulation leads naturally to a set of dynamical diagnostics. The packet centre follows the classical trajectory exactly, while the internal Bohmian deformation is governed by the phase-curvature matrix \(B\) and, at second order, by the curvature-mismatch matrix $\Lambda$. This makes it possible to distinguish bounded rigid transport, quasi-semiclassical deformation, unstable spiral motion, critical runaway behaviour, and non-normalisable sectors within a unified framework. In this sense, Bohmian trajectories do not merely reproduce the classical flow, but provide a direct dynamical probe of coherence, spreading, and instability.

More specifically, the bounded quasi-semiclassical regimes found here provide a Bohmian quantum extension of classical results showing that suitably coupled ghost sectors need not exhibit runaway instabilities \cite{deffayet2022ghosts,deffayet2023global,diez2024foundations}. In the present setting, this persistence of bounded motion is diagnosed through bounded centre dynamics, bounded internal deviation, and the absence of sustained expansion in the internal Bohmian flow.

A compact analytical picture also emerges from the Gaussian ansatz. The deviation \(u=q-q_c\) from the packet centre satisfies the first-order equation
\begin{equation}
\dot u=-GB\,u ,
\end{equation}
and hence the local stability of the Bohmian flow is controlled by the symmetric part of the matrix $-GB$, while the second-order equation
\begin{equation}
\ddot u=-G\Lambda\,u
\end{equation}
encodes the competition between classical and quantum curvature. Bounded Bohmian motion therefore requires three ingredients: a normalisable packet, bounded centre motion, and the absence of sustained expanding directions in the internal flow. These criteria organise the different regimes observed in our numerical analysis.

The main conceptual result of the paper is obtained in the bi-Hamiltonian setting. We showed that two Hamiltonian descriptions which generate the same classical phase-space vector field need not lead to the same Bohmian quantum theory. Although the ghost Hamiltonian \(H_g\) and its classically equivalent partner \(H_2\) produce identical classical trajectories, their Bohmian trajectories and trajectory-evaluated quantum potentials differ because the guidance law depends explicitly on the kinetic tensor of the chosen Hamiltonian representation. Classical equivalence is therefore too weak a criterion for Bohmian quantum equivalence.

This observation is particularly relevant for higher time-derivative systems, where reductions to first-order form and alternative Hamiltonian structures are common. Our results suggest that such formulations should not be compared solely through their classical equations or spectral properties. Trajectory-based diagnostics provide additional information about quantum transport, coherence, and instability, and may therefore serve as a useful tool in assessing competing quantisations of higher-derivative models.

A natural next step would be to extend the present analysis beyond single Gaussian states and to study interference effects, nodal structures, and their impact on Bohmian transport. We leave this question for future work.

\medskip

\noindent {\bf Acknowledgments}: S.D. acknowledges the support of the research grant DST/FFT/NQM/ QSM/2024/3 (by DST-National Quantum Mission, Govt. of India).

\newif\ifabfull\abfulltrue


\end{document}